\documentclass[conference,a4paper]{IEEEtran}
\usepackage{threeparttable}
\usepackage{placeins}
\usepackage{listings}
\usepackage{qtree}
\usepackage{amsmath}
\usepackage{forest}
\usepackage{color}
\usepackage{array}
\usepackage{graphicx}
\usepackage{float}
\usepackage{xspace}
\usepackage[labelfont=bf]{caption}
\newcommand{\nn}{\textit{NN}\xspace}
\newcommand{\knn}{\textit{k}-NN\xspace}
\newcommand{\BigO}[1]{$O\left(#1\right)$\xspace}
\newcommand{\ArrowList}[1]{
	\begin{itemize}
		\item[$\Rightarrow$] #1 
	\end{itemize}
}

\graphicspath{ {graphics/} }

\usepackage{hyperref}

\title{GeoTree: a data structure for constant time geospatial search enabling a real-time mix-adjusted median property price index}

\author{
	\IEEEauthorblockN{
		Robert Miller\IEEEauthorrefmark{1},
		Phil Maguire\IEEEauthorrefmark{2}
	}
	\IEEEauthorblockA{
		Department of Computer Science, \\
		National University of Ireland, Maynooth, \\
		Kildare, Ireland. \\
		Email: \IEEEauthorrefmark{1}robert.miller@mu.ie, \IEEEauthorrefmark{2}phil.maguire@mu.ie
	}
}

\begin{document}
	\maketitle
	
	\begin{abstract}
		A common problem appearing across the field of data science is \knn ($k$-nearest neighbours), particularly within the context of Geographic Information Systems. In this article, we present a novel data structure, the GeoTree, which holds a collection of geohashes (string encodings of GPS co-ordinates). This enables a constant \BigO{1} time search algorithm that returns a set of geohashes surrounding a given geohash in the GeoTree, representing the approximate $k$-nearest neighbours of that geohash. Furthermore, the GeoTree data structure retains an \BigO{n} memory requirement. We apply the data structure to a property price index algorithm focused on price comparison with historical neighbouring sales, demonstrating an enhanced performance. The results show that this data structure allows for the development of a real-time property price index, and can be scaled to larger datasets with ease.
		
	\end{abstract}
	
	\section{Introduction}
	Large scale datasets are a hot topic in computer science. Each one tends to present its own problems and intricacies \cite{Hand2013}. The Nearest Neighbour (\nn) problem is a well known and vital facet of many data mining research topics. This involves finding the nearest data point to a given point under some metric which measures the \textit{distance} between data points. In the context of geospatial data, the \nn{} problem often emerges in the form of geographical proximity search \cite{Roussopoulos:1995:NNQ:568271.223794}. 
	
	Real world geographic data is usually represented by a pair of GPS co-ordinates, which pinpoint any location on Earth with unlimited precision. As a result of their structure, computing the distance between pairs of points in order to find the \textit{nearest neighbour} can be extremely slow on large datasets.
	
	The problem often requires expansion to finding the \textit{k} nearest neighbours (\knn), which further increases the complexity by requiring a sorting of the distance matrix in order to extract a ranking of points by proximity. It is extremely computationally expensive to compute and rank these distances on large datasets \cite{Safar2005}. A computationally cheap method of solving this problem would vastly improve the scalability of proximity based algorithms \cite{Roussopoulos:1995:NNQ:568271.223794}. We propose a data structure which enables such cheap computation, the GeoTree, and explore its potential when applied to a real-world geospatial task.

	\section{Background}	
	\subsection{Naive geospatial search}
	The distance between two pieces of geospatial data defined using the GPS co-ordinate system is computed using the \textit{haversine} formula \cite{10.2307/2309088}.	If we wish to find the closest point in a dataset to any given point in a naive fashion, we must loop over the dataset and compute the haversine distance between each point and the given, fixed point. This is an \BigO{n} computation. If the distances are to be stored for later use, this also requires \BigO{n} memory consumption. Thus, if the closest point to every point in the dataset must be found, this requires an additional nested loop over the dataset, resulting in \BigO{n^2} memory and time complexity overall (assuming the distance matrix is stored). If such a computation is applied to a large dataset, such as the 147,635 property transactions used in the house price index developed by \cite{Maguire2016}, an \BigO{n^2} algorithm can run extremely slowly even on powerful modern machines.
	
	As GPS co-ordinates are multi-dimensional objects, it is difficult to prune and cut data from the search space without performing the haversine computation. With a considerable portion of big data being geospatial in nature, geospatial algorithms and data structures are coming under increased research attention, with the amount of personal location data available growing by approximately 20\% year-on-year according to the \textit{McKinsey Global Institute} \cite{LEE201574}. As such, exploring alternative methods of representing GPS co-ordinates is necessary to make algorithmic improvements.
	
	\subsection{GeoHash}
	
\begin{figure}[h!t]
		\centering
		\includegraphics[width=0.9\linewidth]{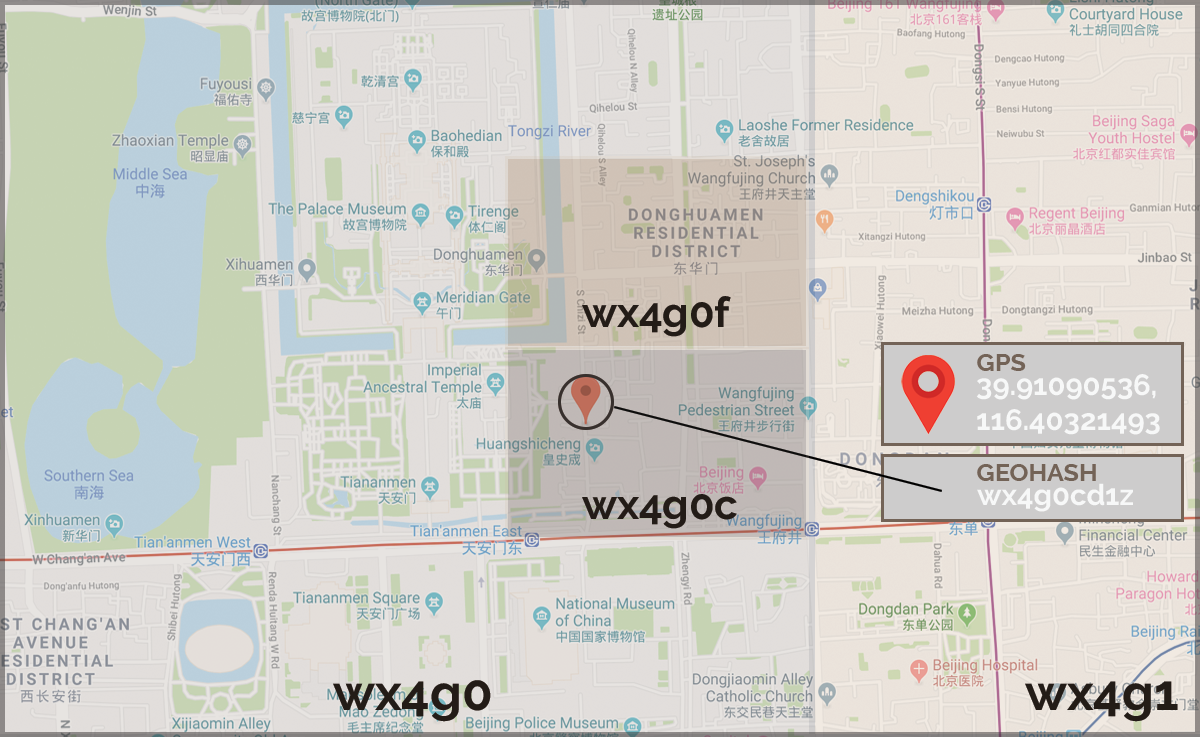}
		\caption{GeoHash algorithm applied to a map}
		\label{fig: geohash_map_demo}
	\end{figure}

	A geohash is a string encoding for GPS co-ordinates, allowing co-ordinate pairs to be represented by a single string of characters. The publicly-released encoding method was invented by Niemeyer in 2008 \cite{GeoHashBlog}. The algorithm works by assigning a geohash string to a square area on the earth, usually referred to as a \textit{bucket}. Every GPS co-ordinate which falls inside that bucket will be assigned that geohash. The number of characters in a geohash is user-specified and determines the size of the bucket. The more characters in the geohash, the smaller the bucket becomes, and the greater precision the geohash can resolve to. While geohashes thus do not represent points on the globe, as there is no limit to the number of characters in a geohash, they can represent an arbitrarily small square on the globe and thus can be reduced to an exact point for practical purposes. \autoref{fig: geohash_map_demo} demonstrates parts of the geohash grid on a section of map.
	
	Geohashes are constructed in such a way that their string similarity signifies something about their proximity on the globe. Take the longest sequential substring of identical characters possible from two geohashes (starting at the first character of each geohash) and call this string $x$. Then $x$ itself is a geohash (ie. a bucket) with a certain area. The longer the length of $x$, the smaller the area of this bucket. Thus $x$ gives an upper bound on the distance between the points. We will refer to this substring as the \textit{smallest common bucket} (SCB) of a pair of geohashes. We define the length of the SCB as the length of the substring defining it. This definition can additionally be generalised to a set of geohashes of any size. Furthermore, we define the SCB of a single geohash $g$ to be the set of all geohashes in the dataset which have $g$ as a prefix. We can immediately assert an upper bound of 123,264m for the distance between the geohashes in \autoref{fig:geo_precision}, as per the table of upper bounds in the \textit{pygeohash} package \cite{McGinnisPygeohashGIT}.
	
		\begin{figure}
			\centering
			\[ \textrm{geohash 1: } \underbrace{c_1c_2c_3}_{\textrm{SCB}}x_4 \dots x_n \]
			\[ \textrm{geohash 2: } \underbrace{c_1c_2c_3}_{\textrm{SCB}}y_4 \dots y_n \]
			
			\noindent
			where: $ x_i \neq y_i \forall i \in \{4 \dots n\} $
			\caption{Geohash precision example}
			\label{fig:geo_precision}
	\end{figure}
	
	\subsection{Efficiency improvement attempts}
	Geohashing algorithms have, over time, improved in efficiency and have been put to use in a wide variety of applications and research contexts \cite{Moussalli2015} \cite{PatentGeohashEnhanced}. As stated by \cite{Roussopoulos:1995:NNQ:568271.223794}, the efficient execution of nearest neighbour computations requires the use of niche spatial data structures which are constructed with the proximity of the data points being a key consideration.
	
	The method proposed by Roussopoulos et al. \cite{Roussopoulos:1995:NNQ:568271.223794} makes use of \textit{R-trees}, a data structure very similar in nature to the geohash \cite{Guttman:1984:RDI:971697.602266}. They propose an efficient algorithm for the precise \nn computation of a spatial point, and extend this to identify the exact $k$-nearest neighbours using a subtree traversal algorithm which demonstrates improved efficiency over the naive search algorithm. Arya et al. \cite{Arya:1998:OAA:293347.293348} further this research by introducing an approximate \knn algorithm with time complexity of \BigO{kd\log n} for any given value of $k$.
	
	A comparison of some data structures for spatial searching and indexing was carried out by \cite{kothuri2002quadtree}, with a specific focus on comparison between the aforementioned \textit{R-trees} and \textit{Quadtrees}, including application to large real-world GIS datasets. The results indicate that the Quadtree is superior to the R-tree in terms of build time due to expensive R-tree clustering. As a trade-off, the R-tree has faster query time. Both of these trees are designed to query for a very precise, user-defined area of geospatial data. As a result they are still quite slow when making a very large number of queries to the tree.
	
	Beygelzimer et al. \cite{Beygelzimer:2006:CTN:1143844.1143857} introduce another new data structure, the cover tree. Here, each level of the tree acts as a "cover" for the level directly beneath it, which allows narrowing of the nearest neighbour search space to logarithmic time in $n$.
	
	Research has also been carried out in reducing the searching overhead when the exact \knn results are not required, and only a spatial region around each of the nearest neighbours is desired. It is often the case that ranged neighbour queries are performed as traditional \knn queries repeated multiple times, which results in a large execution time overhead \cite{BaokRangeNN}. This is an inefficient method, as the lack of precision required in a ranged query can be exploited in order to optimise the search process and increase performance and efficiency, a key feature of the GeoTree.
	
	Muja et al. provide a detailed overview of more recently proposed data structures such as partitioning trees, hashing based \nn structures and graph based \nn structures designed to enable efficient \knn search algorithms \cite{Muja}. The \textit{suffix-tree}, a data structure which is designed to rapidly identify substrings in a string, has also had many incarnations and variations in the literature \cite{SuffixTree}. The GeoTree follows a somewhat similar conceptual idea and applies it to geohashes, allowing very rapid identification of groups of geohashes with shared prefixes. 
	
	The common theme within this existing body of work is the sentiment that methods of speeding up \knn search, particularly upon data of a geospatial nature, require specialised data structures designed specifically for the purpose of proximity searching \cite{Roussopoulos:1995:NNQ:568271.223794}.
	
	\section{GeoTree}
	
	The goal of our data structure is to allow efficient approximate ranged proximity search over a set of geohashes. For example, given a database of house data, we wish to retrieve a collection of houses in a small radius around each house without having to iterate over the entire database. In more general terms, we wish to pool all other strings in a dataset which have a maximal length SCB with respect to any given string.
	
	\subsection{High-level description} \label{sec:high-level-desc}
	
	A GeoTree is a general tree (a tree which has an arbitrary number of children at each node) with an immutable fixed height $h$ set by the user upon creation. Each level of the tree represents a character in the geohash, with the exception of level zero - the root node. For example, at level one, the tree contains a node for every character that occurs among the first characters of each geohash in the database. For each node in the first level, that node will contain children corresponding to each possible character present in the second position of every geohash string in the dataset sharing the same first character as represented by the parent node. The same principle applies from level three to level $h$ of the GeoTree, using the third to $h^{th}$ characters of the geohash respectively. 
	
	At any node, we refer to the path to that node in the tree as the \textit{substring} of that node, and represent it by the string where the $i^{th}$ character corresponds to the letter associated with the node in the path at depth $i$.
	
\begin{figure}[h!t]
		\centering
		\begin{forest}
			for tree={font=\scriptsize,fit=tight,for children={l sep+=0.5em,l+=0.5em}}
			[ROOT [a 
			[a [$\dots$ ]]
			[b [$\dots$ ]]
			[$\dots$ [$\dots$ ]]
			[z [$\dots$ ]]
			]
			[b [$\dots$ ]]
			[$\dots$ [$\dots$ ] ]
			[z [$\dots$ ]]
			]
		\end{forest}
		\caption{GeoTree General Structure}
		\label{fig:gt_struct}
	\end{figure}
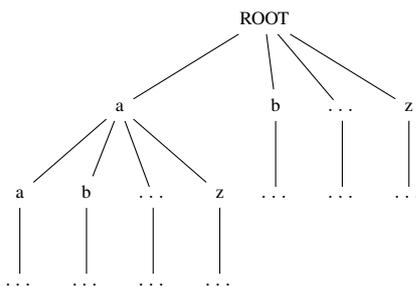

\begin{figure}[h!t]
			\centering
			\begin{forest}
				for tree={font=\scriptsize,fit=tight,for children={l sep-=1em,l-=1em}}
				[ROOT
				[a 
				[c 
				[7
				[j
				[9
				[8
				]
				]
				]
				]
				]
				]
				[g 
				[c 
				[7 
				[j 
				[9 
				[8
				]
				[d 
				]
				]
				]
				]
				[9 
				[a 
				[a 
				[j 
				]
				]
				]
				]
				]
				[d 
				[7 
				[j 
				[9 
				[8 
				]
				]
				[y 
				[a 
				]
				]
				]
				]
				]
				]
				]
			\end{forest}
			\caption{Sample GeoTree Structure}
			\label{fig:gt_sampletree}
	\end{figure}
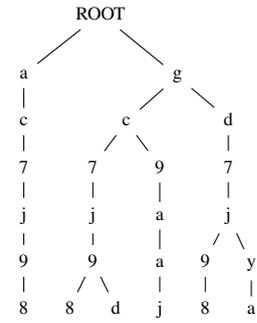

	The general structure of a GeoTree is demonstrated in \autoref{fig:gt_struct}. As can be seen, the first level of the tree has a node for each possible letter in the alphabet. Only characters which are actually present in the first letters of the geohashes in our dataset will receive nodes in the constructed tree. We, however, include all characters in this diagram for clarity. In the second level, the $a$ node also has a child for each possible letter. This same principle applies to the other nodes in the tree. Formally, at the $i^{th}$ level, each node has a child for each of the characters present among the $(i+1)^{th}$ position of the geohash strings which are in the SCB of the current substring of that node. A worked example of a constructed GeoTree follows in \autoref{fig:gt_sampletree}.
	
	Consider the following set of geohashes which has been created for the purpose of demonstration: $\{gc7j98, gc7j98, gd7j98, ac7j98, gc9aaj, gc7j9d, ac7j98, \\
	gd7jya, gc9aaj\}$. The GeoTree generated by the insertion of the geohashes above with a fixed height of six would appear as seen in \autoref{fig:gt_sampletree}.
	
	\subsection{GeoTree Data Nodes}
	
\begin{figure}[h!t]
		\centering
		\captionsetup{justification=centering}
		\begin{forest}
			for tree={font=\scriptsize,fit=tight,for children={s sep-=1em,s-=1em,l sep+=0.1em,l+=0.1em}}
			[ROOT [a 
			[a [$\dots$ 
			[$\{d_1\}$ ]
			[$\{d_2\}$ ]
			]]
			[$\dots$ [$\dots$ 
			[$\{d_3\}$ ]
			]]
			[z [$\dots$ 
			[$\{d_4\}$ ]
			]]
			]
			[$\dots$ [$\dots$ [$\dots$ 
			[$\{d_5\}$ ]
			]]]
			]
		\end{forest}
		\caption{GeoTree Structure with Data Nodes}
		\label{fig:gt_struct_datanodes}
	\end{figure}
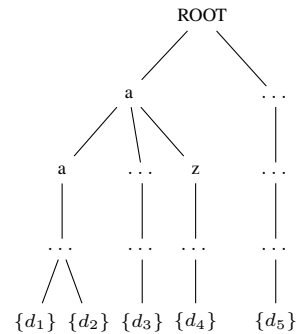

	The data attributes associated with a particular geohash are added as a child of the leaf node of the substring corresponding to that geohash in the tree, as shown in \autoref{fig:gt_struct_datanodes}. In the case where one geohash is associated with multiple data entries, each data entry will have its own node as a child of the geohash substring, as demonstrated in the diagram.
	
	It is now possible to collect all data entries in the SCB of a particular geohash substring without iterating over the entire dataset. Given a particular geohash in the tree, we can move any number of levels up the tree from that geohash's leaf nodes and explore all nearby data entries by traversing the subtree given by taking that node as the root. Thus, to compute the set of geohashes with an SCB of length $m$ or greater with respect to the particular geohash in question, we need only explore the subtree at level $m$ along the path corresponding to that particular geohash. Despite this improvement, we wish to remove the process of traversing the subtree altogether.
	
	\subsection{Subtree Data Caching}
	
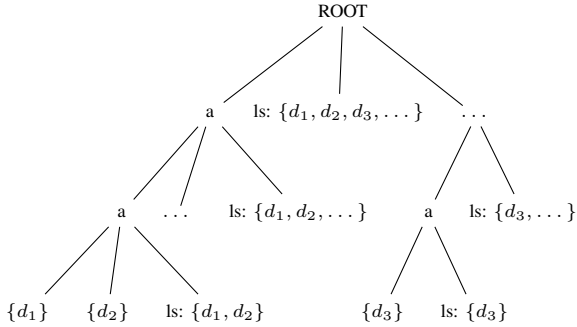
\begin{figure}[h!t]
			\centering
			\begin{forest}
				for tree={font=\scriptsize,fit=tight,for children={l sep+=1em,l+=1em}}
				[ROOT
				[a
				[a
				[{$\{d_1\}$}]
				[{$\{d_2\}$}]
				[{ls: $\{d_1, d_2\}$}]
				]
				[{$\dots$}]
				[{ls: $\{d_1, d_2, \dots\}$}]]
				[{ls: $\{d_1, d_2, d_3, \dots\}$}]
				[{$\dots$}
				[a
				[{$\{d_3\}$}]
				[{ls: $\{d_3\}$}]]
				[{ls: $\{d_3, \dots\}$}]]
				]
			\end{forest}
			\caption{GeoTree Structure with List Nodes}
			\label{fig:gt_struct_listelems}
	\end{figure}

	In order to eliminate traversal of the subtree we must cache all data entries in the subtree at each level. To cache the subtree traversal, each non-leaf node receives an additional child node which we will refer to as the \textit{list} (\textit{ls}) node. The list node holds references to every data entry that has a leaf node within the same subtree as the list node itself. As a result, the list node offers an instant enumeration of every leaf node within the subtree structure in which it sits, removing the need to traverse the subtree and collect the data at the leaf nodes. The structure of the tree with list nodes added is demonstrated in \autoref{fig:gt_struct_listelems} (some nodes and list nodes are omitted for the sake of brevity and clarity).

	\subsection{Retrieval of the Subtree Data}
	
	Given any geohash, we can query the tree for a set of nearby neighbouring geohashes by traversing down the GeoTree along some substring of that geohash. A longer length substring will correspond to a smaller radius in which neighbours will be returned. When the desired level is reached, the cached list node at that level can be queried for instant retrieval of the set of approximate \knn of the geohash in question.
	
	As a result of this structure's design, the GeoTree does not produce a distance measure for the items in the GeoTree. Rather, it clusters groups of nearby data points. While this does not allow for fine tuning of the search radius, it allows a set of data points which are geospatially close to the specified geohash to be retrieved in constant time.
	
	\subsection{Memory Requirement of the Data Structure}
	
	As each geohash is associated with only one character at each level of the GeoTree, only one node on each level will hold that geohash's data entry in its list node. Thus, each data entry is inserted into one single list node at every level of the tree. Given a tree of height $h$, this means that the data will be stored in $h$ different list nodes in addition to the one leaf node which the data receives. If the dataset is of size $n$, then there will be $(h+1)*n$ data entries stored in the tree. However, as the height of the tree is fixed and specified prior to the building of the tree, the overall memory requirement of the GeoTree is \BigO{n}. This can be further improved to only $n$ data entries stored by collecting a set of the data once in memory and filling the list nodes with a list of pointers to the data entries, if necessary.
	
	\subsection{Technical Implementation}
	
	To touch briefly on the implementation of GeoTree \cite{Geotree_RMiller}, a nested hash map structure is used in order to store the tree. The root node is the root hash map of the nest, with the hash keys at this level corresponding to the letters of the level one nodes. Each of these keys point to a value which is another hash map containing keys corresponding to the level two letters of geohashes which have matching first letters with the parent key. The nesting process continues down to the leaf nodes (or terminal hash values in this case) in the same fashion described in \autoref{sec:high-level-desc}. The final hash key (representing the last character of the geohash) points to the list of data entries associated with that geohash. 
	
	\subsection{Time Complexity}
	\subsubsection{Building (Insertion)}
	
	As hash maps offer \BigO{1} insertion, insertion of data at each level of the GeoTree is \BigO{1}. Furthermore, due to the height of the tree, $h$, being constant and fixed, insertion of entries to the GeoTree is an \BigO{1} operation overall.
	
	\subsubsection{SCB Lookup}
	
	The \BigO{1} lookup of hash maps also means that the tree can be traversed in steps of \BigO{1} time. As the \textit{list} nodes hold the SCB of every geohash substring possible from those in the dataset, and a maximum of $h$ SCBs will need to be queried, it follows that any SCB lookup is also \BigO{1}.
	
	\subsection{Comparison with set enumeration trees (SE-trees)}
	
	The SE-tree, or \textit{set enumeration tree}, is a power set data structure which creates a branching tree of all possible subsets of a set of variables \cite{rymon1992search}. The set enumeration tree shares some basic similarities with the GeoTree. There are, however, fundamental differences between these data structures. The \textit{set enumeration tree} is a structure defined on sets which, by definition, do not consider the ordering of variables. While the SE-tree contains all possible subsets of a set of variables, it does not contain all possible ordered collections of those variables. For example, $\{A,B\}$ will be contained in the SE-tree of variables $\{A, B, C\}$, yet $\{B,A\}$ will not appear in the tree.
	
	In the case of the GeoTree, all possible combinations of characters must be considered, as geohashes are sensitive to ordering. The geohash $gh1992a$, for example, corresponds to an entirely different geographical location than $hg1992a$, despite both containing the same characters in slightly different ordering. The GeoTree is designed to support this sensitivity to ordering, whereas the \textit{set enumeration tree} is not. Furthermore, the \textit{set enumeration tree} has no provision for the cached list nodes of data, which is perhaps the most crucial feature of the GeoTree. Although many interesting algorithms for traversing the SE-tree are explored in \cite{rymon1992search}, they are irrelevant in the present context, as the data structure in question is not designed for proximity search but for the purpose of classification.
	
	\section{Real-World Performance} \label{performance_results}
	
	\subsection{Application: House Price Index Algorithm}
	
	In order to test the performance of GeoTree in practice, we applied it to the computation of an Irish house price index. House price indexes and forecasting models have come under increased attention from a data mining context, with a view to improve the current methods of calculating and forecasting property price changes. Such algorithms could help identify price bubbles, facilitating preemptive measures to avoid another market collapse \cite{Klotz2016,Diewert2015,Jadevicius2015}. 
	
	Many of these algorithms are based around the mix-adjusted median or central price tendency model, which requires a geospatial \knn search \cite{Maguire2016,Goh2012}. This approach is based on the principle that large amounts of aggregated data will cancel noise and result in a stable, smooth signal. It also offers the benefit of being less complex than the highly-theoretical hedonic regression model. It also requires less data than the repeat-sales model, in the sense of both quantity and time period spread \cite{Maguire2016,Goh2012,Prasad2008}.
	
	Maguire et al. \cite{Maguire2016} introduced an enhanced central-price tendency model which outperformed the robustness of the hedonic regression method used by the Irish Central Statistics Office \cite{OHanlon2011}. The primary limitation of this method is the algorithmic complexity and brute-force nature of the geospatial search, which impinges on its scalability to larger datasets, and restricts the introduction of further parameters. Our aim was to apply the GeoTree data structure to improve the execution time, scalability and robustness of this method. We re-implemented the algorithm used by \cite{Maguire2016} (described below), running the algorithm on the same data set (Irish Property Price Register) used in the original article as a control test for performance before introducing the GeoTree. For the purposes of algorithmic complexity calculation, we let $n$ be the average number of house sales present in one month of the dataset, and let $t$ be the number of months of data in the dataset. \\

	\noindent
	Stage two (voting) of the \textbf{original} algorithm is executed as follows:

		\ArrowList{Iterate over each month, $m$, of the dataset \\
			($t$ operations)
			\ArrowList{Iterate over each house, $h$, sold during $m$ \\
				($n$ operations)
				\ArrowList{Iterate over houses sold in $m$ to find the nearest to $h$
					($n$ operations*)}
			}	
		}
	
	\vspace{\baselineskip}

	\noindent
	Stage four (stratification) of the \textbf{original} algorithm is executed as follows:

		\ArrowList{Iterate over each month, $m$, of the dataset \\
			($t$ operations)
			\ArrowList{
				Iterate over each house, $h$, soldHHP during $m$ \\
				($n$ operations)
				\ArrowList{
					Iterate over each month prior to $m$, $m_p$ \\
					($\frac{t-1}{2}$ operations)
					\ArrowList{Iterate over houses sold in $m_p$ to find the nearest to $h$ ($n$ operations*)}
				}
			}
		}

	\vspace{\baselineskip}
	By introducing the GeoTree to the algorithm, the steps which formerly required an \BigO{n} iteration over all houses in the dataset to identify the nearest house (marked by an asterisk) now become an \BigO{1} GeoTree ranged proximity search operation. There is, however, a mild trade-off. Rather than returning the closest property to the house in question, the GeoTree structure instead returns everything in a small area around the house (formally, it returns the maximal length non-empty SCB for that house's geohash). The bucket can then be iterated over to find the true closest property, or an alternative strategy can be employed, such as taking the median price of all houses within the small area.
	
	\subsection{Performance Results}
	
\begin{figure*}[h!t]
		\begin{minipage}{\linewidth}
			\captionof{table}{Complexity and performance of the algorithms}
			\centering
			\newcolumntype{C}[1]{%
				>{\vbox to 3ex\bgroup\vfill\centering\arraybackslash}%
				p{#1}%
				<{\vskip-\baselineskip\vfill\egroup}} 
			\newcolumntype{?}[1]{!{\vrule width #1}}
			\resizebox{0.9\linewidth}{!}{
				\begin{threeparttable}
					\begin{tabular}{ | C{6.5em} ?{0.3em} C{7.5em} ?{0.3em} C{7.5em} | C{4em} ?{0.3em} C{7.5em} | C{4em} | }
						\hline Algorithm & Complexity & $\mu$ (1 core)\tnote{\scriptsize a} & $\sigma$\tnote{b} & $\mu$ (8 cores)\tnote{\scriptsize a} & $\sigma$\tnote{b} \\
						\hline \hline
						Voting & {\large \BigO{n^2t} } & 233.54 seconds\tnote{c} & 2.37\% & 46.73 seconds\tnote{c} & 1.69\% \\
						\hline
						\textbf{Voting\textsuperscript{+}} & {\large \BigO{nt} } & 12.78 seconds\tnote{c} & 1.68\% & 3.02 seconds\tnote{c} & 0.69\% \\
						\hline \hline
						Stratify & {\large \BigO{\frac{n^2t(t-1)}{2}} } & 29.03 hours & 2.41\% & 4.19 hours & 1.89\% \\
						\hline
						\textbf{Stratify\textsuperscript{+}} & {\large \BigO{\frac{nt(t-1)}{2}} } & $\mathtt{\sim}$0.05 hours (163.89s) & 1.71\% & $\mathtt{\sim}$0.01 hours (39.63s) & 0.85\% \\
						\hline \hline
						Overall & {\large \BigO{\frac{n^2t(t+1)}{2}} } & 29.11 hours & 2.43\% & 4.21 hours & 1.90\% \\
						\hline
						\textbf{Overall\textsuperscript{+}} & {\large \BigO{\frac{nt(t+1)}{2}} } & $\mathtt{\sim}$0.05 hours (177.73s) & 1.67\% & $\mathtt{\sim}$0.01 hours (43.71s) & 0.79\% \\
						\hline
					\end{tabular}
					\vspace*{0.2em}
					\begin{tablenotes}
						\item[a]	Execution times reported are the mean ($\mu$) of ten trials.
						\item[b]	Standard deviation ($\sigma$) reported as a percentage of the mean ($\mu$).
						\item[c]	\textbf{Includes build time} for the dataset array / GeoTree on the dataset, as applicable.
						\item[d]	All algorithms computed using an AMD Ryzen 2700X CPU.	
						\item[e] 	All algorithms executed on the Irish Residential Property Price Register database of \textbf{279,474 property sale records} as of time of execution.	
					\end{tablenotes}
			\end{threeparttable}}
			\label{tab: gt_complexitytime_orig}
		\end{minipage}

		\vspace*{3em}

		\begin{minipage}{\linewidth}
		\centering
		\includegraphics[width=\linewidth]{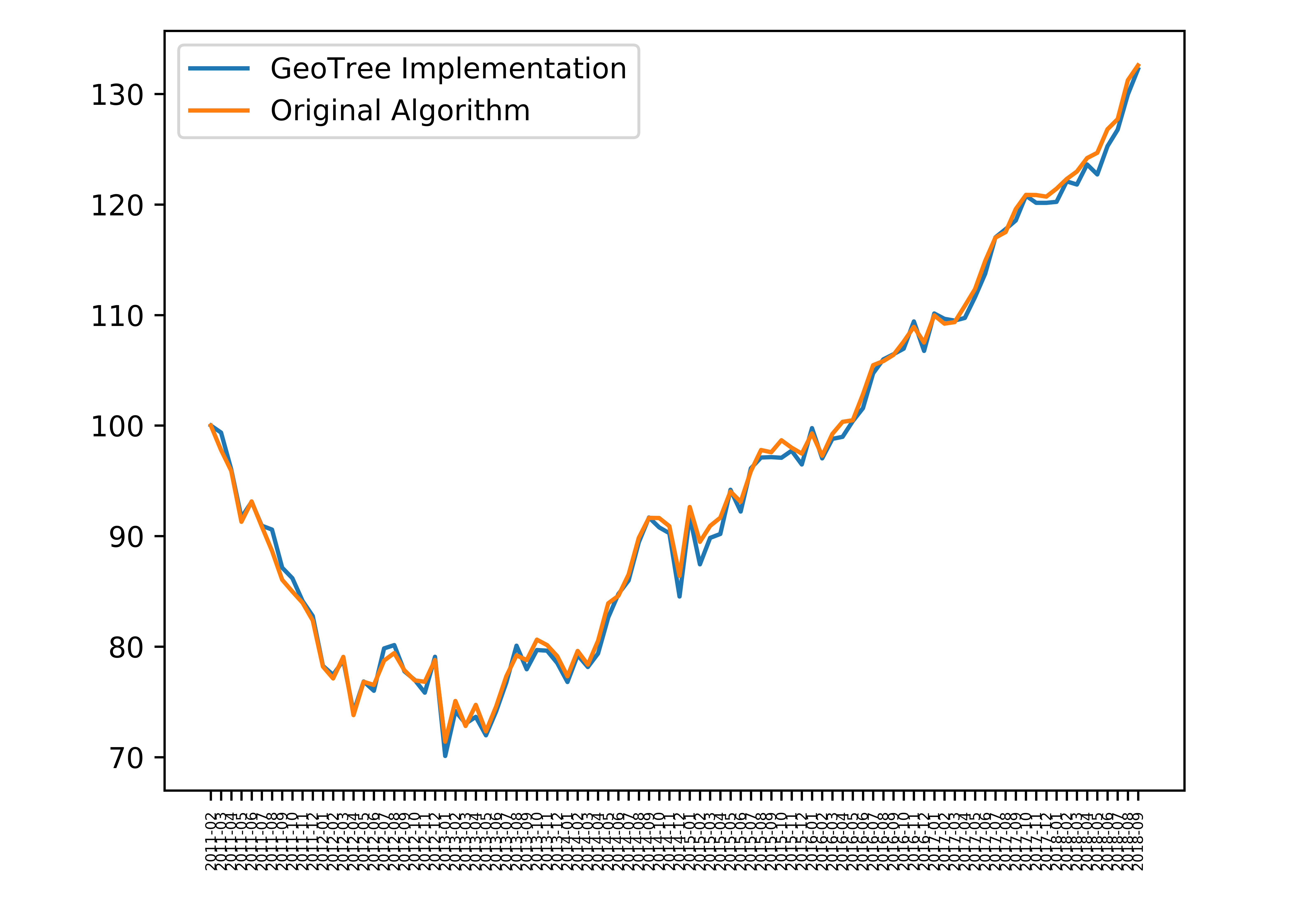}
		\caption{Irish RPPI (GeoTree vs Original), from 02-2011 to 09-2018}
		\label{fig: ppr_graph_geohashcomp}
		\end{minipage}
	\end{figure*}

	\autoref{tab: gt_complexitytime_orig} compares the performance of the algorithms described previously with and without GeoTrees (on a database of 279,474 property sale records), including both single threaded execution time and multi-threaded execution time (running eight threads across eight CPU cores) on our test machine. The results using the GeoTree are marked with a \textit{+} symbol.

	\subsection{Correlation}
	
	Despite the algorithmic alteration of taking the median price of a group of geohashed nearest neighbours, as opposed to the nearest neighbour per se, the house price indexes produced by the original algorithm and the GeoTree-enhanced version are very similar. \autoref{fig: ppr_graph_geohashcomp} shows both versions of the Residential Property Price Index (RPPI) superimposed. The two different versions yielded highly correlated outputs (Pearson's $r$ = 0.999, Spearman's $\rho$ = 0.997, Kendall's $\tau$ = 0.966), revealing that GeoTree succeeded in delivering an almost identical index to the original, though with major performance gains in execution time.
	
	\subsection{Scalability Testing}
	
	In order to test the scalability of the GeoTree, we obtained a dataset comprising 2,857,669 property sale records for California, and evaluated both the build and query time of the data structure. \autoref{tab:scale_gt_perf} shows mean build time and mean query time on both 10\% ($\sim$285,000 records) and 100\% ($\sim$2.85 million records) of the dataset. In this context, query time refers to the total time to perform \textbf{100 sequential queries}, as a single query was too fast to accurately measure.
	
	The results demonstrate that the height of the tree has a modest effect on the build time, while dataset size has a linear effect on build time, thus supporting the claimed \BigO{n} build time with \BigO{1} insertion. Furthermore, query time is shown to remain constant regardless of both tree height and dataset size, with negligible differences in all instances.
	
	\begin{table}
		\centering
		\caption{Scalability Performance of GeoTree}
		\newcolumntype{C}[1]{%
			>{\vbox to 3.3ex\bgroup\vfill\centering\arraybackslash}%
			p{#1}%
			<{\vskip-\baselineskip\vfill\egroup}} 
		\newcolumntype{?}[1]{!{\vrule width #1}}
		\resizebox{\columnwidth}{!}{
		\begin{threeparttable}
			\centering	
			\begin{tabular}{ | C{5.5em} ?{0.25em} C{4.75em} | C{4.75em} | C{4.75em} | C{4.75em} | C{4.75em} | }
				\hline Height $h$ & 4 & 5 & 6 & 7 & 8 \\
				\hline \hline
				Build Time (10\%)\tnote{a} & \textbf{17.63s} (0.08s)  & \textbf{18.10s} (0.10s) & \textbf{18.46s} (0.22s) & \textbf{18.84s} (0.08s) & \textbf{19.39s} (0.09s) \\
				\hline
				Build Time (100\%)\tnote{b} & \textbf{179.67s} (0.58s)  & \textbf{183.80s} (0.57s) & \textbf{183.99s} (0.52s) & \textbf{192.06s} (0.60s) & \textbf{194.31s} (0.94s) \\
				\hline
				Query Time (10\%)\tnote{c} & \textbf{5.1ms} (0.3ms) & \textbf{5.2ms} (0.4ms) & \textbf{5.3ms} (0.9ms) & \textbf{5.3ms} (0.4ms) & \textbf{5.3ms} (0.5ms) \\
				\hline
				Query Time (100\%)\tnote{c} & \textbf{5.4ms} (1.0ms) & \textbf{5.3ms} (0.9ms) & \textbf{5.5ms} (1.0ms) & \textbf{5.7ms} (1.3ms) & \textbf{5.6ms} (1.2ms) \\
				\hline
			\end{tabular}
			\vspace*{0.5em}
			\begin{tablenotes}
				\item[a] Build Time (10\%) is the total time to insert 10\% of dataset ($\mathtt{\sim}$285,000 records)
				\item[b] Build Time (100\%) is the total time to insert 100\% of dataset ($\mathtt{\sim}$2.85m records)
				\item[c] Query Time consists of total time to execute 100 sequential neighbour queries on 10\% and 100\% of the dataset respectively
				\item[d] Times reported are in the format $\boldsymbol{\mu} (\sigma)$ calculated over ten trials
			\end{tablenotes}
			\vspace{-1em}
		\end{threeparttable}}
		\label{tab:scale_gt_perf}
	\end{table}
	
	\section{Conclusion}
	
	We have shown that the GeoTree data structure introduced in this article offers an efficient \BigO{1} method for geospatial approximate \knn search over a collection of geohashes. The application to a real-world property price index algorithm revealed significant reductions in execution time, and potentially opens the door for a real-time property price index. The data structure also performed well when applied to a much larger dataset, demonstrating its scalability. In conclusion, any data science problem which requires geospatial sampling around a particular area can employ the GeoTree for \BigO{1} retrieval of approximate neighbours, potentially enabling, for example, fast retrieval of locations of interest to map users, or geo-targeted advertisement and social networking updates.
	
	\bibliographystyle{IEEEtran}
	\bibliography{IEEEabrv,references}
\end{document}